\documentclass[10pt,conference]{IEEEtran}
\IEEEoverridecommandlockouts
\usepackage[letterpaper, left=0.7in, right=0.7in, top=0.76in, bottom=1.1in]{geometry}

\usepackage{cite}
\usepackage{amsmath,amssymb,amsfonts}
\usepackage{algorithmic}
\usepackage{algorithm,algorithmic}
\usepackage{graphicx}
\usepackage{textcomp}
\usepackage{xcolor}
\usepackage{booktabs}
\def\BibTeX{{\rm B\kern-.05em{\sc i\kern-.025em b}\kern-.08em
    T\kern-.1667em\lower.7ex\hbox{E}\kern-.125emX}}

\makeatletter

\let\old@ps@IEEEtitlepagestyle\ps@IEEEtitlepagestyle

\def\confheader#1{%
    \def\ps@IEEEtitlepagestyle{%
        \old@ps@IEEEtitlepagestyle%
        \def\@oddhead{\strut\hfill#1\hfill\strut}%
        \def\@evenhead{\strut\hfill#1\hfill\strut}%
    }%
    \ps@headings%
}
    
\makeatother
    \confheader{%
    \parbox{20cm}{Accepted by Fourth Workshop on Machine Learning and Deep Learning for Wireless Security, IEEE Global Communications \\ Conference 2025 (IEEE GLOBECOM), \textcopyright 2025 IEEE}
}

\begin{document}

\title{A Digital Twin Platform for QoS Optimization Under DoS Attacks for Next Generation Radio Networks
\thanks{}
}

\author{
\IEEEauthorblockN{Mehmet Ali Ertürk\IEEEauthorrefmark{1}\IEEEauthorrefmark{2}, Kubra Duran\IEEEauthorrefmark{2}, Ahmed Al-Dubai\IEEEauthorrefmark{2} and Berk Canberk\IEEEauthorrefmark{2}}

    \IEEEauthorblockA{\IEEEauthorrefmark{1}Computer Engineering Department, Istanbul University, Istanbul, Turkey
    \\\{mehmetali.erturk\}@istanbul.edu.tr}

    \IEEEauthorblockA{\IEEEauthorrefmark{2}School of Computing, Engineering and The Built Environment, Edinburgh Napier University, UK
    \\\{m.erturk, kubra.duran, a.al-dubai, b.canberk\}@napier.ac.uk}        
}

\maketitle
\begin{abstract}
Digital Twins are being used as an enabling technology in 6G applications across various domains, valued for their data-driven insights and real-time decision-making capabilities. However, integrating Digital Twins into 6G environments presents challenges in maintaining consistent network services under adverse conditions such as including denial-of-service (DoS) attacks, while ensuring consistent Quality of Service (QoS). In this work, we present a Digital Twin Platform to facilitate bidirectional communication between User Equipment (UEs) and application-specific digital twins to enhance UE traffic under UDP flood attacks. By leveraging AI to analyze key digital twin parameters—such as throughput and delay, our framework derives actionable insights that enhance QoS management in DoS attack scenarios, ultimately advancing real-world applications of digital twins in critical infrastructure domains. The performance of this Digital Twin Platform is validated through an emergency management use-case in 6G networks while the network is under attack with UDP flood attacks in terms of packet reception success rate, average packet delay, and average throughput metrics.
\end{abstract}

\begin{IEEEkeywords}
6G, digital twin, quality of service, denial-of-service, cybersecurity
\end{IEEEkeywords}

\section{Introduction}

Future networking relies on highly reliable communication infrastructures to support fully autonomous and rapid operations to serve divergent Quality of Service (QoS) demands. At this point, Digital Twins are emerging as a key enabler for meeting these QoS demands for their real-time monitoring, dynamic optimisation, and intelligent decision-making capabilities \cite{survey1, 9696282}. On the other hand, 5G networks are still vulnerable to Denial of Service (DoS) attacks, where an attacker can disrupt the network core and lead to the unavailability of network resources \cite{Lalropuia2020, sec3}. Furthermore, any delay or unavailability of the network resources to the UE can be disastrous in a critical environment. In these circumstances, it is necessary to have dynamic, autonomous network management to ensure QoS guarantees under various conditions, such as DoS attacks \cite{sec2}. 

In the current situation, a fully autonomous QoS management scheme is not yet feasible due to the limitations of current 5G networks. The networks can be configured based on specific contexts, including mobility and higher-level service requirements \cite{9369564} with multiple numerologies, allowing for variable subcarrier spacing. Even though 6G communication is still under development and no formal specifications have yet been established, it holds the promise of providing a reliable communication infrastructure, even in chaotic environments, ensuring QoS-aware emergency communications through Artificial Intelligence (AI) driven networks \cite{Islam2024, tgcn}. However, the successive end-to-end implementation of the QoS-awareness requires enhanced what-if testing by observing the real-time dynamics of the networks \cite{bc}. Therefore, at this point, the integration of Digital Twins will enable enhanced testing of AI-driven QoS management approaches before deploying them on the real networks.

As 6G networks are anticipated to employ AI for fully autonomous network management, we propose an AI-driven solution that utilizes Digital Twins to manage 6G networks. This approach focuses on QoS considerations for various application scenarios, with the aim of enhancing network performance while ensuring QoS-awareness. To the best of our knowledge, this is the first study to introduce reactive numerology management in 6G networks using Digital Twins. We evaluate the proposed algorithm using our digital twin platform in conjunction with the Simu5G simulator, which incorporates 5G NR (New Radio) networks, focusing on an emergency use case\cite{simultech20}. Our contributions are as follows:

\begin{itemize}
    \item We provide guaranteed QoS through AI-driven communication for 6G digital twin services during UDP flood attacks.
    \item We use simulated annealing, heuristic search and tree-based boosting algorithms to determine the optimal numerologies for UE carriers for delay minimisation. 
    \item We introduce a scenario-based real-time what-if analysis for 6G networks to fully automate network management for QoS.
\end{itemize}

\begin{table*}[!t] 
\centering
\caption{Proposed Framework and Current State of the Art Studies Comparison}
\label{tab:lit}
\begin{tabular}{l|cccccc}
\toprule
\hline
\multicolumn{1}{c|}{\textbf{Literature}} & \textbf{\begin{tabular}[c]{@{}c@{}}QoS-aware\end{tabular}} & \textbf{\begin{tabular}[c]{@{}c@{}}Numerology reconfig.\end{tabular}} & \textbf{\begin{tabular}[c]{@{}c@{}}Parameter\end{tabular}} & \textbf{\begin{tabular}[c]{@{}c@{}}AI-based\end{tabular}} & \textbf{\begin{tabular}[c]{@{}c@{}}Method\end{tabular}} & \textbf{\begin{tabular}[c]{@{}c@{}}DT integration\end{tabular}} \\ \hline
\cite{qosnumerology}, \cite{sec1}                       & \checkmark   & \checkmark        & throughput             & -                  & heuristic                            & -                    \\ 
\cite{ml}, \cite{qosDrama}                                  & \checkmark    & \checkmark        & packet loss                 & \checkmark                      & RF, DRL                             & -                        \\ 
\cite{scheduleRL}, \cite{adlen}                               & \checkmark    & \checkmark        & goodput, latency                     & \checkmark                & DRL                           &-                         \\
\cite{Yagmur2024}, \cite{dt3}, \cite{gentwin}, \cite{qcsm}     & \checkmark     &- & response time              &\checkmark                         & DRL                           & \checkmark                       \\
Our work                                                      & \checkmark     & \checkmark        &  throughput, latency           & \checkmark              & XGBoost                    & \checkmark                \\ \hline
\bottomrule
\end{tabular}
\end{table*}

The remainder of this paper is organized as follows: Section \ref{sec:related} discusses the current state of the art studies from the literature. Section \ref{sec:model} introduces the proposed AI-based Digital Twin model for QoS management. The experimental study and results are described in Section \ref{sec:eval}. Finally, Section \ref{sec:conclusion} concludes the paper.

\section{Related Works}
\label{sec:related}

The 6G wireless communication technology is expected to exceed 5G in data rates, latency, high-density connectivity, and bandwidth. 
As given in Table \ref{tab:6g-req}, QoS metrics will be the key to assessing network performance and delivering high-fidelity network connectivity in 6G \cite{Ahmed2023}. Regarding these metrics, there are several efforts to efficiently manage 6G service management based on their QoS demands under several attack scenarios. For example, \cite{qosnumerology} introduces a QoS-aware radio resource management framework for multi-numerology 5G systems. It dynamically allocates the spectrum among different numerologies to maximize system throughput. On the other hand, \cite{sec1} proposes a digital twin framework based on ontology to detect distributed DoS (DDoS) while observing the deviation of QoS. In this way, the system becomes more robust against DDoS attacks by taking the advantage of complex relations via ontologies. 

Besides the traditional methods, AI-based solutions are also proposed for numerology reconfiguration under attack scenarios. For instance,  \cite{ml} utilizes the Random Forest (RF) algorithm to decide particular numerologies for different QoS levels. Furthermore, \cite{qosDrama} introduces QoS-DRAMA, a Deep Reinforcement Learning (DRL)-based adaptive resource allocation scheme in 5G and 6G networks. It optimizes scheduling heuristics by dynamically adjusting the weights of multiple scheduling metrics by reducing packet loss rates. Besides, several DRL models are proposed for radio resource scheduling under different numerology settings for 5G services, especially focusing on goodput and latency metrics \cite{scheduleRL, adlen}. Even though all these efforts are to meet with QoS requirements while serving numerology configuration, none of these solutions address the optimality while designing their proposals.



\begin{table}[!ht]
\centering
\caption{Requirements of 6G \cite{Ahmed2023}. }
\label{tab:6g-req}
\begin{tabular}{|l|l|}
\hline
\textbf{Data Rate (Peak)}           & \textgreater 1 Tb/s  \\ \hline
\textbf{User Experienced Data Rate} & \textgreater 10 GB/s \\ \hline
\textbf{Latency}                    & \textless 1m         \\ \hline
\textbf{Max Channel Bandwidth}      & 100 GHz              \\ \hline
\textbf{Connection Density}      & 10 mil. / square km  \\
\hline
\end{tabular}
\end{table}


Future networking requires lower delay tolerance and real-time communications. This requirement paves the way for the development of Digital Twins to meet 6G communication needs, owing to their real-time capabilities. Regarding this, several research studies on Digital Twins have been made to provide reliable, resource-efficient and timely communication. For instance, \cite{Yagmur2024} discusses how to solve the dynamic scheduling problem using Digital Twins and intelligent gate control methods, which ensure highly reliable communications in next-generation smart cities. Besides, \cite{dt3} designs a Digital Twin-based resource allocation framework for dynamic vehicular networks by using a multi-agent deep reinforcement learning approach. Similarly, \cite{gentwin} proposes a Digital Twin framework supported by Generative AI to lower the response time while creating the QoS-aware twin models. Moreover, \cite{qcsm} proposes a Q-learning-based cognitive service management scheme to address QoS degradation in 6G IoT networks. All these efforts show that the Digital Twin approach can efficiently manage low-latency communications in network management, especially with its What-if capability. However, none of these DT efforts address numerology reconfiguration; they perform mainly on the application layer rather than performing physical layer adjustments. Consequently, the integration of Digital Twin-based approaches with AI-based numerology reconfiguration remains a research gap by holding significant potential for meeting the 6G QoS metrics. We give a detailed comparison of these mentioned studies in Table \ref{tab:lit}.



\begin{figure*}[!ht]
    \centering
    \includegraphics[width=0.95\textwidth]{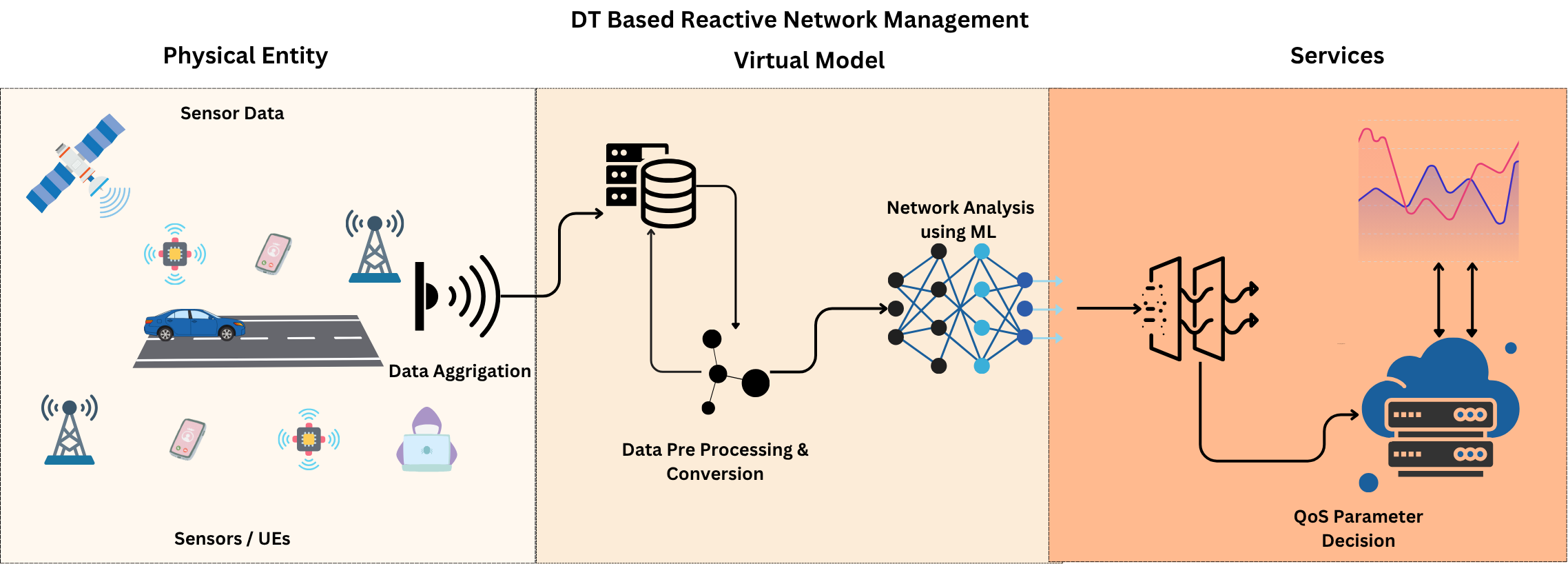}
    \caption{QoS-aware 6G Digital Twin approach for QoS optimization.}
    \label{fig:qos6goverview}
\end{figure*}

\section{Proposed Model}
\label{sec:model}

A 6G network may consist of various communication elements, including satellites, vehicles, mobile phones, and Internet of Things (IoT) sensors. Ensuring the quality of communication links is a challenging task due to the dynamic nature of network topology. Additionally, the communication requirements of User Equipment (UE) can change depending on the context of the application. Also, the network serves as an attack surface where different adversaries can intervene and disrupt the active services. Using digital twins, we propose an AI-based QoS management model that meets the real-time requirement under UDP flood attacks. Figure \ref{fig:qos6goverview} gives an overview of the proposed network solution. 

Digital twins are the virtual representation of physical entities, humans, or processes. It enables real-time communication through two-way data transfer. Data collected from the physical peer can be mirrored in the virtual environment while the virtual model is continuously updated with real-time data. This ability allows for the simulation and analysis of physical entities in a virtual environment, offering insights that surpass traditional simulation methods. 

Our proposed solution utilizes highly accurate digital twin replicas in real-time to analyze network traffic and adaptively manage QoS parameters based on dynamic network conditions. In this study, we will focus on managing QoS during emergency situations. Accordingly, our digital twin is configured to respond to changes in network traffic, specifically during these emergencies. The presented model focuses on 6G networks, where each UE is configured with the necessary network radio interfaces. The algorithm operates as follows:

\begin{itemize}
    
\item For a selected region, the network topology is pre-configured and virtually replicated to indicate the initial point.
\item Each client is configured with the same QoS parameters as a starting point.
\item A client is configured to act as an adversary and generates burst UDP packets to consume bandwidth to disrupt services. 
\item The network performance is monitored in real time in terms of average delay and throughput. 
\item Within DT a what-if test scenario is triggered to have some UEs having emergency data transmission case.
\item UEs QoS configuration is updated through Algorithm \ref{alg:graphconstr} to minimize mean delay in the network in real-time. 
\item The configured parameters obtained for numerologies are reconfigured with DT on the network to evaluate the performance of the network. 
\item The network is monitored in real-time for any emergency situations. 
\end{itemize}

We assume that user space distribution is known to DT, and when a critical event is triggered, the nodes having neighbors can be identified within a simple graph traverse with depth one. We also use the ability of Simu5G to reconfigure numerologies, where NR can adopt various values for respective transmission time intervals (TTI) as outlined in the Table \ref{tab:nr-numerology}.

\begin{table}[!ht]
\centering
\caption{New Radio numerologies \cite{Nardini2020Simu5g}. }
\label{tab:nr-numerology}
\begin{tabular}{|l|l|l|l|l|l|}
\hline
\textbf{Numerology ($\mu$) } & 0 & 1 & 2 & 3 & 4   \\
\hline
\textbf{TTI  duration (ms)} & 1 & 0.5 & 0.25 & 0.125 & 0.625  \\

\hline
\end{tabular}
\end{table}

We utilize different transmission slots to manage network status and meet QoS requirements, as discussed in the study \cite{simultech20}. We also consider each UE's Carrier Component $CC \in \{1, 2, 3, 4, 5\} $ has configurable parameters $\mu \in  \{ 0, 1, 2, 3, 4\} $  that sets transmission slots for the $CC_i$ for the UE. At this juncture, finding the optimum configuration is exponential since its complexity can be represented as $\mathcal{O}({CC}^{\mu})$. Here, our goal is to minimize the average delay in the chosen network region through a simulated annealing heuristic search \cite{AIBook} under UDP flood. For this, we assess each selected parameter using XGBoost and historical network data to reduce the delay experienced by end-users (denoted as $D_{UE}$) by prioritizing packet transmissions as given in (Eq. \ref{eqn1}). We aim to determine the optimal value of $\mu_{j}$ for $CC_i$, which will help decrease network delay. 

\begin{equation}
    \operatorname*{argmin}_{CC_1,\mu_1 ...CC_n,\mu_n} 1/n \sum D_{UE}
    \label{eqn1}
\end{equation}

\begin{minipage}{.9\linewidth}
 \begin{algorithm}[H]
 	 \caption{Proposed algorithm for optimum NR numerology config for delay minimization.}
 	 \label{alg:graphconstr}
 	 \begin{algorithmic}[1]

  \STATE $CC=CC_0$
  \FOR {$m = 0$ to $n\mu$}

   \STATE $netDelay=XGBoostRegMinDelay()$
   \STATE $UE_k=GetRandomNeighbour(UE_i)$

   \STATE $UE_k =  UpdateUENumerology(UE_k, CC, m) $

    \IF { $XGBoostRegMinDelay() < netDelay$ }
        \STATE $CC = CC_{new}$     
    \ENDIF
    
  \ENDFOR
 \RETURN CC
 \end{algorithmic}
 
 \end{algorithm}
\end{minipage}

\section{Performance Evaluation}
\label{sec:eval}

In this section, we investigate the performance of the proposed AI-based Digital Twin model for QoS optimization and management. For this, we measure (i) packet reception success rate, (ii) average packet delay (in $seconds$), and (iii) average throughput (in $bps$) values for an increasing number of UEs. We evaluate single cell and multi cell scenarios and compare the results for default numerology and DT-managed numerology.

\begin{table}[!h]
    \centering
\caption{Simulation parameters} 
\centering 
\begin{tabular}{l c} 
\hline 
Parameters&Values \\ [0.5ex]
\hline\hline 
Simulation area & 1000$x$1000 $m^2$ \\
Number of UEs &  \{10, 25, 50, 100, 200, 250\} \\
Number of carriers & $[1-5]$\\
$F_c$ & [2-6]\\
Learning rate & 0.1\\
Confidence interval  & 95\% \\
\hline 
\end{tabular}
\label{tab:sim}
\end{table}

\begin{figure}[!ht]
    \centering
    \includegraphics[width=0.50\textwidth]{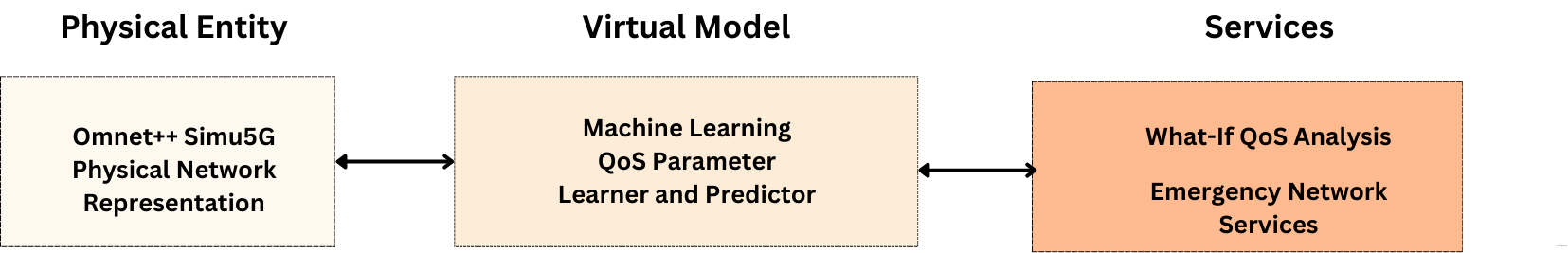}
    \caption{Evaluation flow for proposed Digital Twin-based QoS optimization framework.}
    \label{fig:graphcomplex}
\end{figure}

\begin{figure}[!ht]
    \centering
    \includegraphics[width=0.50\textwidth]{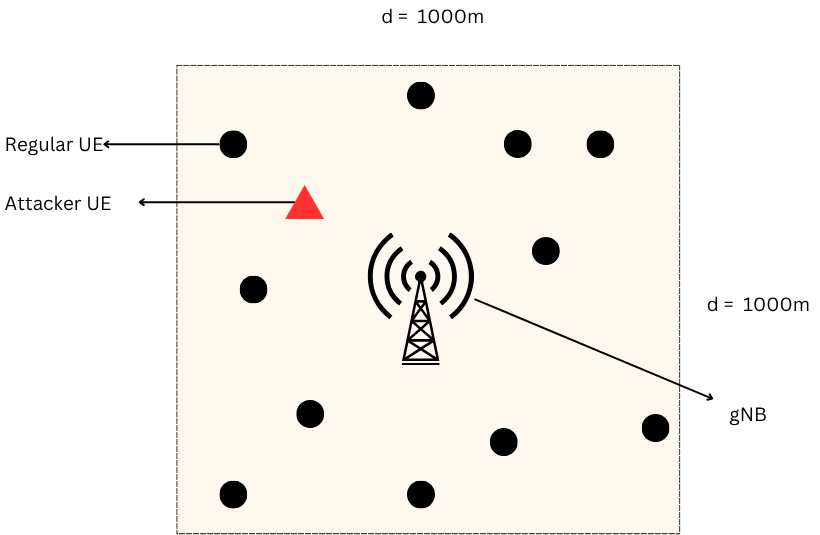}
    \caption{Evaluation network overview.}
    \label{fig:simarea}
\end{figure}

We first designed a physical network scenario using Simu5G, a library in Omnet++ that focuses on evaluating 5G networks \cite{Nardini2020Simu5g}. 
Afterwards, we created the virtual model and then ran the what-if for QoS analysis. The flow of our evaluation step is given in Figure \ref{fig:graphcomplex}. We also give the simulation parameters in Table \ref{tab:sim}. In our experimental scenario, UEs are equipped with NR interfaces. The number of end users is defined by the $ U = \{10, 25, 50, 100, 200, 250 \}$, and they are randomly distributed within a grid area. Each UE is assigned to a gNbs, as shown in Figure \ref{fig:simarea}. Each UE is configured with a different number of carriers, \( C = \{1, 2, 3, 4, 5\} \), and operates on carrier frequencies \( F_c = [2 - 6] \). In this configuration, the simulation environment represents the physical network connected to our Digital Twin platform. The network is configured to have a saturation condition in which the UE always has a packet to transmit in the queue. The following scenarios are evaluated within the experimental study.

\begin{itemize}
    \item Default NR configuration for the single cell scenario. 
    \item Default NR configuration for multi-cell scenario. 
    \item DT managed NR optimization for the single cell scenario. 
    \item DT managed NR optimization for multi-cell scenario. 
    
\end{itemize}

Network parameters are captured in real-time to monitor network traffic in terms of global network delay and throughput, as well as per UE.  Through our DT environment, we are able to perform a what-if analysis where simulating an emergency condition is a breakout at event t within selected random UEs. The DT is capable of predicting network throughput and delay through historical data. Each UE with NR is reconfigured through multiple numerologies that allow our DT to manage QoS requirements for the selected emergency service scenarios.

Network performance prediction is managed using gradient boosting regression algorithms, specifically XGBoost. Tree boosting has gained attention in machine learning applications due to its strong predictive performance. It can be applied effectively to standard classification and regression problems. XGBoost is an efficient machine learning system designed for tree boosting \cite{Chen2016XGBoost}. As an open-source tool, XGBoost is capable of handling large datasets, even in memory-limited environments. This makes it an effective candidate for real-time analysis.

Our goal is to minimize the average delay in the critical region of the network during emergencies, as data transmission is vital. To achieve this, we will search for optimal transmission slots that reduce average delays across the region while reconfiguring all UEs in the network. The parameters are searched heuristically by using the simulated annealing probabilistic method. In each iteration, the critical region delay is minimized to find the optimum minimum for the scenario. 

Real-time data is processed through the XGBoost regression and evaluated based on the following features:

\begin{itemize}
    \item Number of gNB (Next Generation Node B)
\item Number of connected User Equipments (UE)
\item Number of carriers 
\item Packet size 
\item UE carrier count 
\item UE carrier bandwidth
\item UE numerology 
\item UE delay  
\item UE throughput 
\item UE successful packet transmission count 
\item UE dropped packet count 

\end{itemize}

This evaluation helps to predict network performance accurately.

\begin{figure}[!ht]
    \centering
    \includegraphics[width=0.47\textwidth]{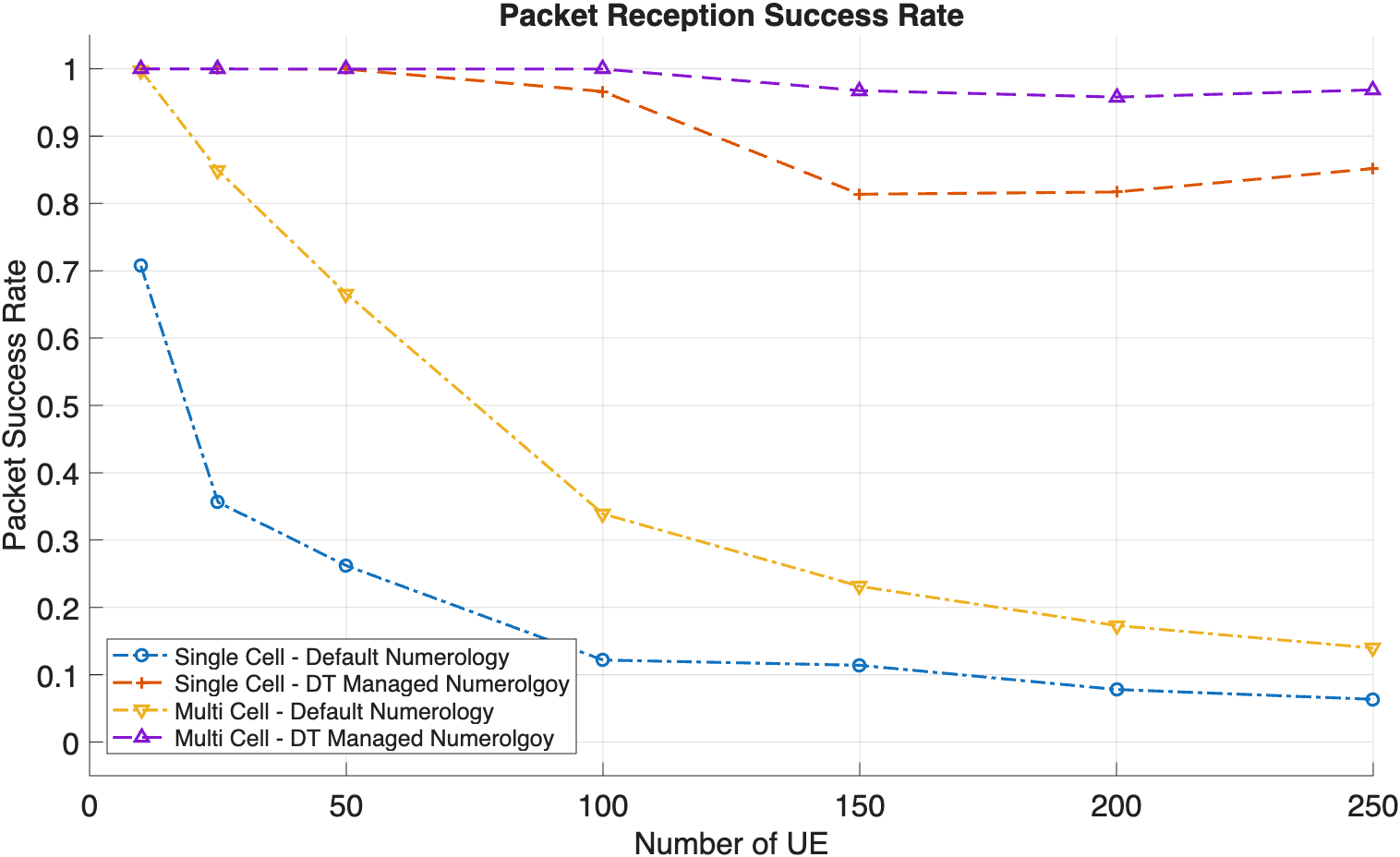}
    \caption{Comparison of packet success ratio for different scenarios.}
    \label{fig:packetsucc}
\end{figure}

\begin{figure}[!ht]
    \centering
    \includegraphics[width=0.47\textwidth]{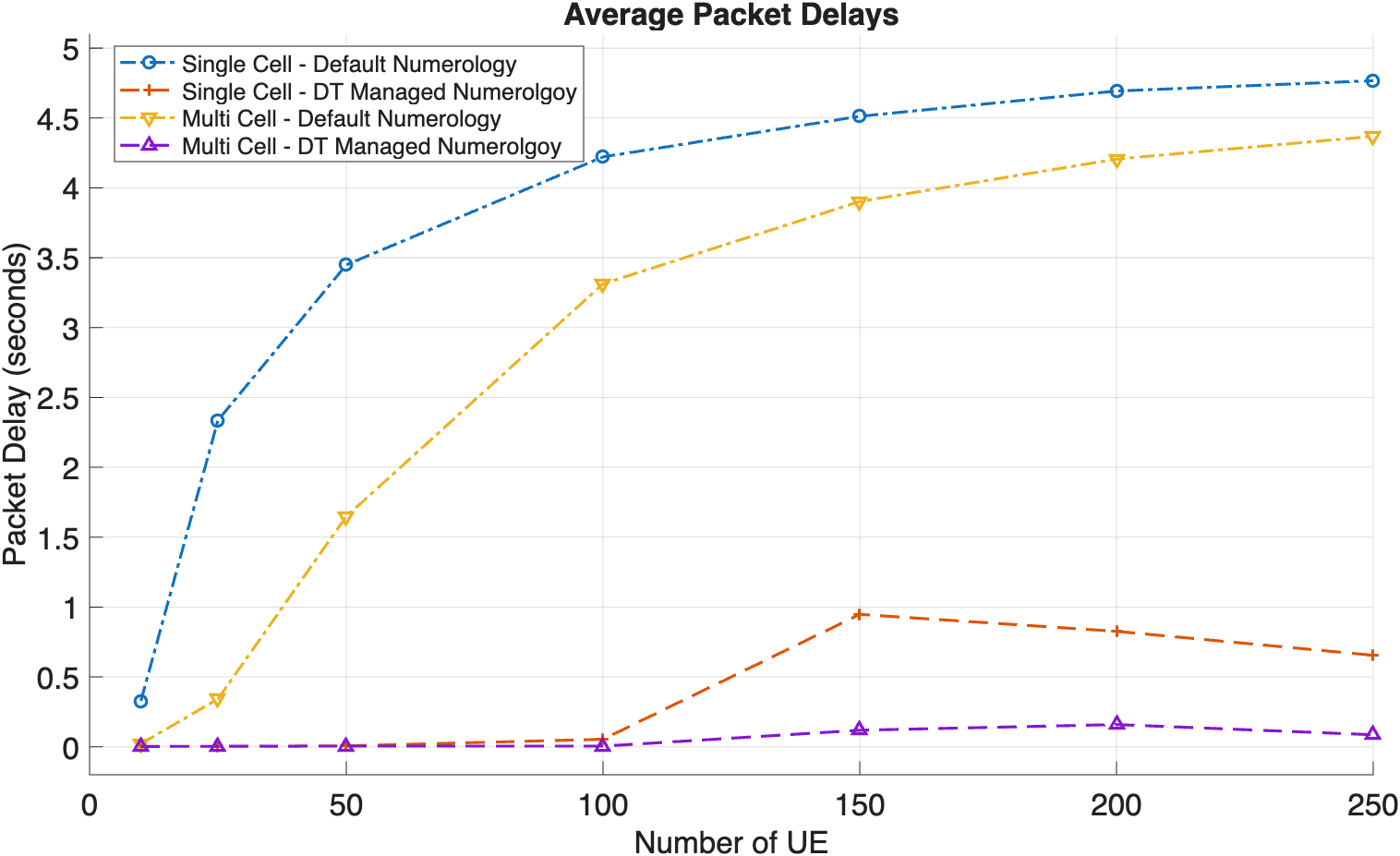}
    \caption{Comparison of average delays across different scenarios.}
    \label{fig:avgdelay}
\end{figure}

\begin{figure}[!ht]
    \centering
    \includegraphics[width=0.47\textwidth]{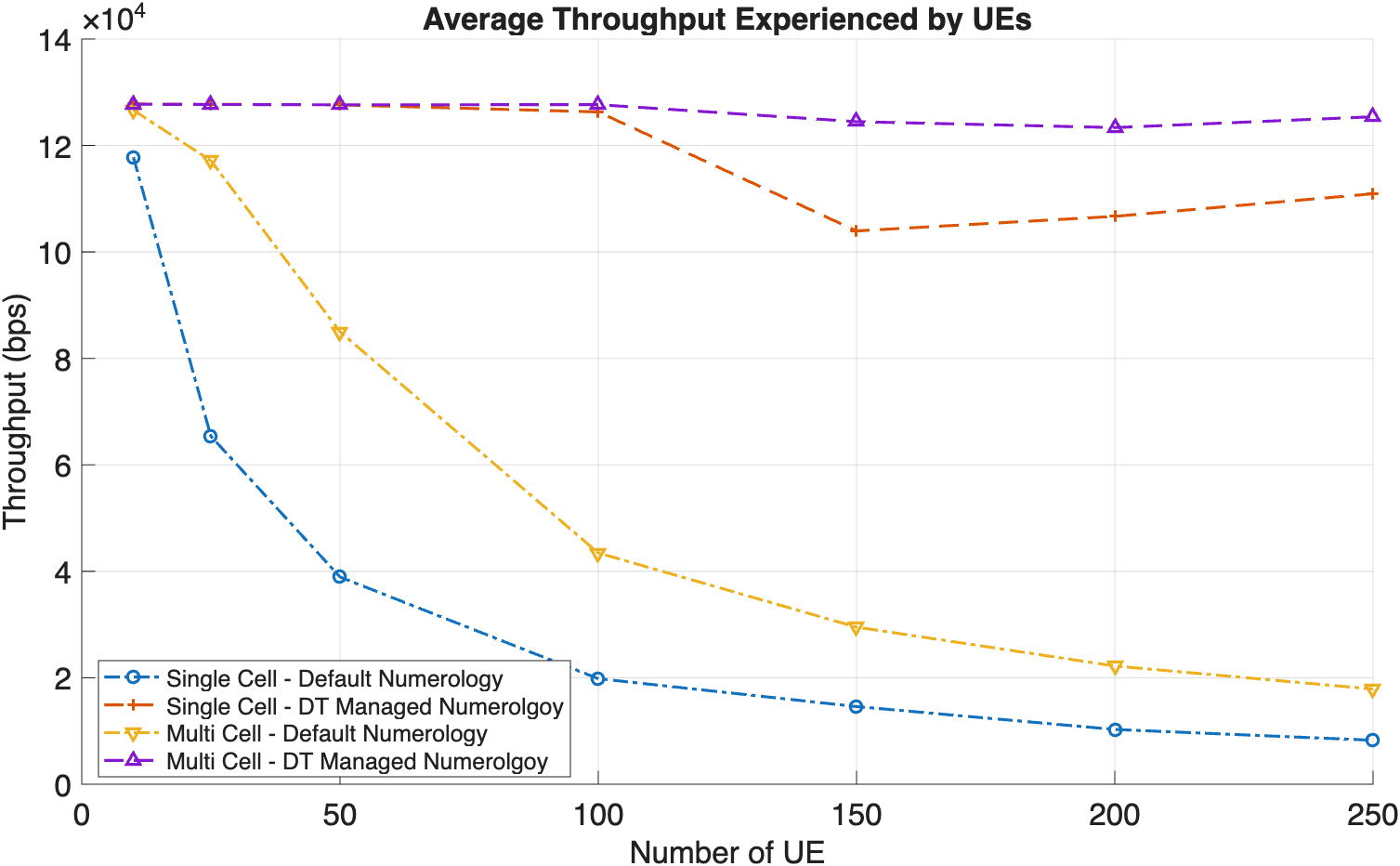}
    \caption{Comparison of average throughput across different scenarios.}
    \label{fig:throughput}
\end{figure}

The experimental results are presented in Figures \ref{fig:packetsucc} to \ref{fig:throughput}. In Figure \ref{fig:packetsucc}, a successful packet reception ratio is given. Notably, default numerology configuration cases degrade quickly when the number of UE increases in the network. However, when we optimize the network numerology parameters, performance increases quickly, even in a higher number of UEs. Also, average packet delays in the network behave in a similar pattern. In Figure \ref{fig:avgdelay}, the average delay packet transmission delay experienced by UEs is given. The default configuration exhibits worse delay rates than the DT-managed solution in single and multi-cell network configurations.  It is clear that the existing default numerology selection strategy has longer waiting times during frame transmission compared to our proposed solution. Furthermore, Figure \ref{fig:throughput} demonstrates the average UE throughput (Bps) for different configurations. Since throughput is calculated with successfully transmitted packets, a similar pattern exists with the results in Figure \ref{fig:packetsucc}.

Overall, the simulation results show that when combined with Digital Twin systems, AI-based QoS management is more effective than traditional methods, especially as the number of UEs increases. Therefore, it’s important to note that using Digital Twin-based optimization in a multi-cell scenario ensures QoS guarantees for critical applications, such as transportation systems or disaster management scenarios which require emergency communication.

\section{Conclusion and Future Work}
\label{sec:conclusion}

Next-generation 6G communication networks are designed to operate with fully autonomous functionalities, enhancing services for mission-critical applications. These 6G network services can be managed in real-time using digital twin systems, enabling completely autonomous management with the help of AI. In this study, we present an AI-based Quality of Service (QoS) management framework using digital twins for optimising the network under cyber attacks. We focus on emergency communications on 6G networks to reduce network delays experienced by users. The results indicate that the autonomous management of New Radio (NR) numerology can effectively decrease network delay during critical events through real-time configuration. 

In future work, we plan to extend our proposed framework for dynamic network prediction to use in extreme 6G mobility cases. We plan to integrate this with real-time digital twin systems. 
\section*{Acknowledgment}

This work was partially supported by the Scientific and Technological Research Council of Turkey (TUBITAK) 1515 Frontier R\&D Laboratories Support Program for BTS Advanced AI Hub: BTS Autonomous Networks and Data Innovation Lab. Project 5239903. The work of M. A. Erturk was supported by the Scientific and Technological Research Council of Turkey (TUBITAK) 2219- International Postdoctoral Research Fellowship Program (App. No: 1059B192301494). 




\bibliographystyle{IEEEtran}
\bibliography{bib}

\end{document}